\documentclass[aps,pre,twocolumn,nofootinbib]{revtex4-1}
\usepackage{amsmath,amssymb,amstext,graphicx}
\begin{document}
\title{Physics from Angular Projection of Rectangular Grids}
\author{Ashmeet Singh}
\email{ashmtuph@iitr.ac.in}
\affiliation{Department of Physics, Indian Institute of Technology Roorkee, India - 247667}
\begin{abstract}
In this paper, we present a mathematical model for the angular projection of a rectangular arrangement of points in a grid. This simple, yet interesting problem, has both a scholarly value and applications for data extraction techniques to study the physics of various systems. Our work can interest undergraduate students to understand subtle points in the angular projection of a grid and describes various quantities of interest in the projection with completeness and sufficient rigour. We show that for certain angular ranges, the projection has non-distinctness, and calculate the details of such angles, and correspondingly, the number of distinct points and the total projected length. We focus on interesting trends obtained for the projected length of the grid elements and present a simple application of the model to determine the geometry of an unknown grid whose spatial extensions are known, using measurement of the grid projection at two angles only. Towards the end, our model is shown to have potential applications in various branches of physical sciences including crystallography, astrophysics and bulk properties of materials.
\end{abstract}
\maketitle
\section{Introduction} \label{intro}
Many physical systems have order in the arrangement of their constituent elements -  the study of which enables us to understand the structure and various properties of the system. A sound understanding of this internal order can help us devise relevant data extraction techniques which harness prior information about the order to help probe the system better. A typical example of utilizing the order for the study of the system can be directly seen in, but is in no way limited to, crystallography\cite{periodic, kittle}, such as Bravais lattices\cite{bravais}, where physical properties of the system such as the material density, electronic properties, Young's modulus, Bulk modulus, coefficient of thermal expansion etc. are dependent on the crystal structure (order) of the lattice. 

In this paper, we study a simplistic arrangement of a rectangular grid of constituent elements; which is a common, two dimensional geometrical arrangement seen in many physical scenarios. The grid elements are arranged in a rectangular fashion with constant spacing between neighbouring elements and this constant spacing or the period, may have different values along the two perpendicular directions of the rectangular grid. We mathematically model the angular projection of the grid elements onto a circular screen and try to understand the angular dependence of the projection pattern on various grid parameters. In particular, we study the variation of the projected length, the number and distribution of projected points with the angle of projection, and try to correlate the modeled trends with the grid parameters. The constructed model will be shown to have practical implementations in physics; as data extraction techniques -  which combine a physical model of the structure with observations to make inferences about the system properties. As we will show, the formulated model can be used to determine the number of grid elements, the size of each element and spacing between them in an unknown grid, using only two angular projection measurements. With a simple setup coupled with tools and methodology available to young students, they can appreciate both the reflective and application-oriented nature of this work.

In section \ref{grid}, we describe the grid setup, describing various parameters in detail and explain the idea of the angular projection of the grid. In section \ref{model}, we build the mathematical model of the angular projection and calculate quantities of physical interest which is extended in section \ref{two_angle}, where we discuss a methodology based on the model to determine the grid parameters using a two-angle measurement approach. In Section \ref{applications}, we outline a number of possible applications of our work and its possible extensions in various branches of physical sciences such as crystallography, astrophysics, signal encoding and bulk properties of mechanical systems.  We conclude in section \ref{conclude} by summarizing our work.
\section{The Grid Setup} \label{grid}
\begin{figure}
  \includegraphics[scale = 0.31]{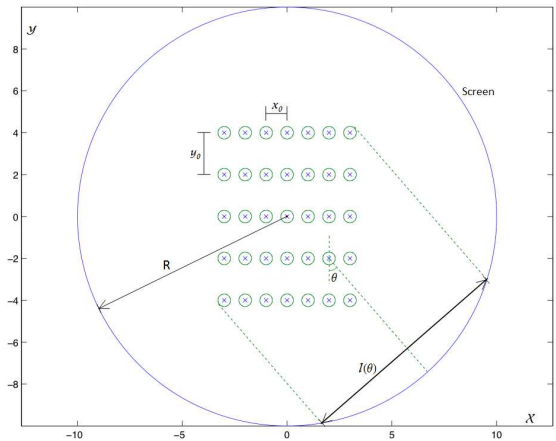}
  \caption{An Example Setup of the Rectangular Grid of 7 $\times$ 5 points, each associated with a finite sized circular element (size exaggerated for clarity \textit{i.e.} not to scale). Also shown are the spacing between neighbouring grid centers, a general projection angle, total projected chord length and the circular screen.}\label{gridsetup}
\end{figure}

Consider a rectangular grid of points having $n$ rows and $m$ columns, that is a total number of $N_{0} = n m$ grid points. Each grid point has an associated circular grid element of radius $a_{0}$. For convenience, we have taken both $n$ and $m$ to be odd positive integers and have chosen an orthogonal Cartesian coordinate system having axes $x$ and $y$, and the origin is coincident with the center point of the grid. Immediately, we see that the grid is symmetric with respect to both $x$ and $y$-axes. Let $x_{0}$ and $y_{0}$ be the constant spacing between neighbouring grid point centers along the $x$ and $y$ directions, respectively. The grid setup is shown in Figure (\ref{gridsetup}). We use discrete Latin indices $(i,j)$ to label the grid point center in the $i$-th column from the right and $j$-th row from the bottom (ref. Figure \ref{gridsetup}) such that a general point $P$ on the grid can be labeled as,
\begin{equation}
\label{pointP}
P  \equiv P_{ij} = P(x_{i} , y_{j}) = \left( \frac{x_{0}}{2} (m - 2i +1) , \frac{y_{0}}{2} (n - 2j +1) \right)	\:	,
\end{equation} 
with $i \in \left\lbrace 1,2, 3,..., m  \right\rbrace$ and $j \in \left\lbrace 1,2, 3,..., n \right\rbrace$.
The grid is, therefore, completely characterized by $n,m,x_{0}$, $y_{0}$ and $a_{0}$, which we call the parameters of the grid. The total grid length along the $x$ and $y$-directions is, respectively, $L_{x}$ and $L_{y}$:
\label{lxly}
\begin{equation}
L_{x} = (m-1) x_{0} 		\: \: \:  ; \: \: \:  L_{y} = (n-1) y_{0}	 \: 
\end{equation}
The angular projection of the grid at an angle $\theta$ (as shown in Figure (\ref{gridsetup})) is the projection of all $N_{0}$ grid elements onto a circular screen of radius $R$, also having its center at the origin of the coordinate system. The angle $\theta$ is measured from the negative $y$-axis with $\theta \in \left[ 0, {2\pi} \right] $ and is conventionally taken to be positive in the counter-clockwise direction. We exclusively work in the range $\theta \in \left[ 0, \frac{\pi} {2}\right] $, since for higher angles, the extension of the results is straightforward due to the two-fold symmetry in the grid setup. This projection can be physically realized by passing a light beam (with a plane wave-front)\cite{hecht} through the grid at different angles, and detecting it on the concentric circular screen screen kept in the $x-y$ plane. The shadow regions, against an illuminated background on the screen will be the projection of the grid elements. These projections can be detected using a photo-detector\cite{photodetector} which can move along the screen to measure the changes in intensity of light due to the grid pattern. With the grid setup so ready, we now model the angular projection and relate it to the grid parameters.
\section{Modeling the Angular Projection} \label{model}
\begin{figure}
  \centering
  \includegraphics[scale = 0.33]{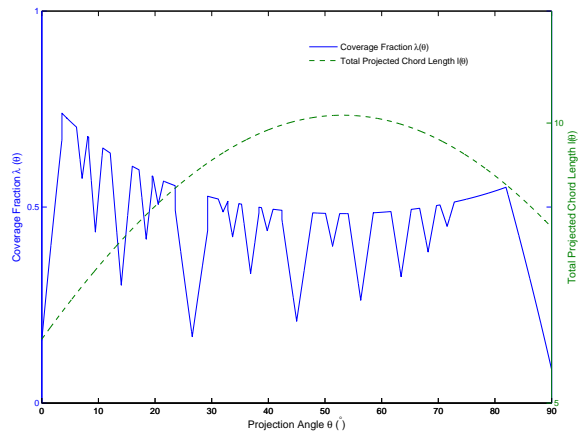}
  \caption{Coverage Fraction(solid line) and Total Projected Chord length (dashed line) of the Angular Projection of a Grid having $n = 5, m=7$ and $x_{0} = 1$, $y_{0} =2$, $R = 10$ and $a_{0} = 0.07$ in standard distance units.}\label{thefig1}
\end{figure}

We now model the angular projections of the grid onto the concentric circular screen and develop ways to utilize this model in physical situations for appropriate data extraction applications. We first consider the extent of the projection on the circular screen, that is, the length between the two most separated (projected) points. It is readily seen that this length $l(\theta)$ will be the projected length of the grid diagonal. We first get an expression for the projected length of the intercepting chord, which can then be used to find the arc length of the projection along the screen. The chord length can be understood as the projection of the line joining the diagonally opposite grid centers and an additional projected length of $2 a_{0}$ due to the finite radii of the grid elements located at the diagonally opposite ends,
\begin{equation}
\label{ltheta_chord}
l(\theta) = L_{x} \cos\theta + L_{y} \sin\theta + 2a_{0} 	\:	.
\end{equation} 
This can be seen to have a maximum value for $\theta = \tan^{-1}(L_{y}/L_{x})$. The length of the projected arc along the circular screen will be denoted be $\widehat{l(\theta)}$ and can be found using the fact that the angle subtended by the projection at the grid center is $\left[ 2 \sin^{-1} (l(\theta) / 2R) \right]$, where $R$ is the radius of the screen, such that,
\begin{equation}
\label{ltheta_nm}
\widehat{l(\theta)} = 2R \sin^{-1} \left( \frac{l(\theta)}{2R} \right) 	\:	.
\end{equation}
Naturally, in the angular range $[0,2\pi]$ the projected arc length of the grid is an oscillatory function of the projection angle $\theta$. In our range $[0,\pi/2]$, the projected chord length of the grid is shown as the dotted line in Figure \ref{thefig1} which corresponds to a grid having $n = 5, m=7$ with $x_{0} = 1 , y_{0}=2$, $R = 10$ and $a_{0} = 0.07$ in standard distance units.
We now go on to determine the angular dependence of the number of distinct projected grid points, the projected length due to the grid elements and other related parameters useful for a data extraction model. At any general $\theta$, we expect a maximum number of $N_{0}$ projections of the grid elements, each having a one-dimensional projected length of $2a_{0}$ on the projection chord. In certain angular ranges, however, the projection of some grid elements may overlap with one another, and in such a case, each grid element will not project itself distinctly. We call this the degenerate scenario, since multiple elements on the grid will have a finite overlap with other grid elements. At certain specific values of $\theta$, there would be a complete overlap between different grid elements and at such angles, the number of projected grid elements will decrease. Around each such specific angle, there would be a finite angular width, within which even though the grid \textit{centers} will be distinct, there will be overlap between different grid elements due to the finite extent ($\sim a_{0}$) of each element. But the projection at a non-degenerate angle will have all the $N_{0} = nm$ grid elements (and therefore, the grid centers) projected distinctly, each spanning $2a_{0}$ in length. Let $L_{P} (\theta)$ be the projected length of the chord due to the grid elements at angle $\theta$. For non-degenerate angles, the projection due to the grid elements will have a length given by,
\begin{equation}
\label{L_P_nondegenerate}
L_{P} (\theta) = 2nm a_{0} \: , \: \: \: \theta \textsl{ is a non-degenerate angle} \: .
\end{equation}
However, this condition of non-degeneracy must be accompanied by a constraint on the size of each grid element. We need to ensure that at non-degenerate angles, there is no overlap between grid elements due to a large $a_{0}$ since all element projections are confined to a finite chord length $l(\theta)$. We impose the following strict condition on the size of each grid element in view of this,
\begin{equation}
\label{condition}
(L_{P} (\theta) = 2nm a_{0}) < \textsf{min} (l(\theta)) \: .
\end{equation}
This condition essentially demands that the projected length of the grid elements $L_{P} (\theta)$ should not exceed the total expanse of the grid projection $l(\theta)$. We make it strict by imposing that the maximum possible length of the grid element projections must be smaller than the smallest value of $l(\theta)$, ensuring there would be no overlap at any non-degenerate angle due to the finite size of each element. Though it must be pointed out that the minimum of $l(\theta)$ could be either  $L_{x} +2a_{0} \:  \textsl{or} \:  L_{y} + 2a_{0}$, and correspondingly  $L_{P} = 2ma_{0}$ or $2na_{0}$, respectively, and not $2nma_{0}$. Thus, the condition of Eq. (\ref{condition}) presents a very strict limit on $a_{0}$ which can be succinctly written as,
\begin{equation}
\label{condition2}
a_{0} < \frac{L_{x} + L_{y}}{4(nm - 1)} 	\:	\:	.
\end{equation}
Please note that the grid element size $a_{0}$ is exaggerated in Figure \ref{gridsetup} for visual clarity. The grid element radius $a_{0}$ must follow Eq. (\ref{condition2}). The next crucial aspect is to understand the degeneracy in the grid projection due to the angular orientation of the projection \textit{i.e.} when their projections start overlapping at certain angles of projection. 
 Let $N(\theta)$ be the number of projections of distinct grid element centers at angle $\theta$. It is easily seen that for all those angles such that $\theta \neq \tan^{-1}({p x_{0}}/{q y_{0}})$, with any combination of $p,q$ satisfying,
\begin{eqnarray} 
\label{pqRestrict}
p \in \left\lbrace 0,1,2,...,(m-1) \right\rbrace 		\: , \\  \nonumber
q \in \left\lbrace 0,1,2,...,(n-1) \right\rbrace	 \: , \\ \nonumber
p = q = 0 \textsl{ is not allowed}	\:,
\end{eqnarray}
the number of distinct centers is $N(\theta) = N_{0} = nm$, since for these angles, all the grid centers will have a distinct projection.  We label such angles by the set $\left\lbrace \alpha \right\rbrace$, such that whenever $\theta \in \left\lbrace \alpha \right\rbrace$, each grid center has a distinct projection. Let, $\chi = \tan^{-1}({p x_{0}}/{q y_{0}})$ be a subset of all possible angles $\theta$, labeled by the set $\left\lbrace \chi \right\rbrace$, for which all grid element centers do not have a distinct projection. Thus, whenever $\theta \in \left\lbrace \chi \right\rbrace$, there exist degeneracies in the projection of the grid and this is termed as the ``full-eclipse" scenario, since the alignment between the grid elements is perfect. In the full-eclipse case, we have, as expected, $N(\chi) < N_{0}$. The number of such angles in the set $\left\lbrace \chi \right\rbrace$ is the number of unique combinations of $({p}/{q})$ in the lowest rational form satisfying Eq. (\ref{pqRestrict}). As an illustration of this point, we see that $p=0$ for any $q \neq 0$ corresponds to $\chi \equiv \theta = 0$, in which case, there are only $m$ distinct projections of the grid centers, one due to each column. On the other extreme, $q = 0$ for any $p \neq 0$, will correspond to $\chi \equiv \theta = {\pi}/{2}$, and there would be only $n$ distinct projected centers, one due to each row. We label each angle in the set $\left\lbrace \chi \right\rbrace$ with a subscript $``k"$ such that a full-eclipse angle is labeled as $\chi_{k}$. At $\chi_{k}$, many grid elements will be degenerate behind a single grid element and there will be a total of $N(\chi _{k})$ number of distinct grid elements projected on the screen. Around each $\chi_{k}$ , there will be an associated angular width $\Delta \chi_{k}$ associated with it, within which there will be overlap between projection of grid elements, even though the element centers are distinct. In the range $\left\lbrace \chi_{k} \pm \Delta \chi_{k} \right\rbrace$ \textit{i.e.} when $\theta \in \left[\chi_{k} - \Delta\chi_{k} , \chi_{k} + \Delta\chi_{k} \right]$, the element centers will be distinct but the projection of the grid elements will overlap such that each of the $N(\chi_{k})$ elements which were distinct at $\chi_{k}$ will now be ``smeared out" due to the finite size of each element. 
We first calculate the number of distinct grid centers projected at a given degenerate angle $\chi_{k}$ and the projected length of the grid elements, before moving on to the more involved case when $\theta \in \left\lbrace\chi_{k} \pm \Delta \chi_{k} \right\rbrace$. Only rational combinations of $({p}/{q})$ in the lowest form satisfying Eq. (\ref{pqRestrict}) will contribute to a unique $\chi$. To ensure that we work only with unique angles $\chi$ while considering all combinations of Eq. (\ref{pqRestrict}), we reduce $({p}/{q})$ to its lowest rational form and write it as $({u}/{v})$, such that $u$ and $v$ are co-prime. Thus, there is a many-to-one map between the pair $(p,q)$ to the $(u,v)$ pair, such that each $(p,q)$ combination can be mapped to a corresponding $(u,v)$ pair, and hence, to the angle $\chi = \tan^{-1}({u x_{0}}/{v y_{0}})$. It is worth mentioning here that the lowest rational combination for $p=0$ for any $q \neq 0$ corresponds to the $(0,1)$ $(u,v$ pair and for $q=0$ for any $p \neq 0$, to the pair $(1,0)$. 
 Let, for a given $(u,v)$ pair, the total number of points (grid centers) which have a non-distinct projection be $\beta (u,v) \equiv \beta(\chi_{k})$, such that the number of distinct projected points will be,
\begin{equation}
\label{distinctPoints}
{N} (\chi_{k}) = N_{0} - \beta (\chi_{k}) = nm - \beta (\chi_{k})		\:	.
\end{equation}
As mentioned in section \ref{grid}, we label the row of grid points closest to the bottom of the grid as ``\textit{Row-1}'', the next as ``\textit{Row-2}'' and so on till ``\textit{Row-$n$}''. We analyze the rows for the projection of their points in this numerical sequence to determine the total number of distinct projections. The number of points of \textit{Row-$t$} having a non-distinct projection due to overlap with projections of other points in the same row or points of previously analyzed rows is given by $\beta _{t} (u,v)$ such that the total number of such degenerate (non-distinct) points in the grid is,
\begin{equation}
\label{beta1}
\beta (u,v) = \sum_{t = 1}^{n} \beta _{t} (u,v)	\:	.
\end{equation}
It can be seen with some thought that for \textit{Row-1}, there are $(m-u)$ points having a non-distinct projection if and only if $v = 0$ (which corresponds to $u=1$ and $\chi = \frac{\pi}{2}$), else all points will have a distinct projection. Thus, when $(u,v) = (1,0)$, we see that $(m-1)$ points of \textit{Row-1} have a degenerate projection, and there is only $u=1$ distinct projected point for this row. From \textit{Row-2}, there would be $u$ points having distinct projections, so that $(m-u)$ of its $m$ points have a non-distinct projection only if $v=1$ as they would overlap with the last $(m-u)$ points of \textit{Row-1}. We emphasize that the readers give it sufficient thought to convince themselves that for \textit{Row-$t$}, there would again be $u$ distinct projections and correspondingly $(m-u)$ of its points having non-distinct projections only if $v \in \left\lbrace 0,1,2,3,...,t-1 \right\rbrace$. The value of $v$ determines with which row the $(m-u)$ points of \textit{Row-$t$} will overlap, such that $v = 0$ corresponds to the row itself, $v = 1$ signifies overlap with the points of \textit{Row-$(t-1)$} and so on. This can be expressed by formulating an expression for $\beta _{t} (u,v)$ with the use of Kronecker delta\cite{kronecker} functions as follows,
\begin{equation}
\label{betat}
\beta _{t} (u,v) = (m-u) \sum_{k = 0}^{t-1} \delta_{v,k}	\:	.
\end{equation} 
With an expression for $\beta _{t}(u,v)$ ready, we can easily find the total number of grid points which have a non-distinct projection using Eq. (\ref{beta1}),
\begin{equation}
\label{beta2}
\beta (u,v) = \sum_{t=1}^{n} \sum_{b = 0}^{t-1}  (m-u) \delta _{v,b}	\:	,
\end{equation}
which can be further simplified as, 
\begin{equation}
\label{beta4}
\beta (u,v) =  (m-u) \sum_{s=0}^{n-1} (n-s) \delta _{v,s} 	\:	.
\end{equation}
For any given $\chi$, $v \in \left\lbrace 0,1,2,...,(n-1) \right\rbrace$ from Eq. (\ref{pqRestrict}), depending on the values of $p,q$ and $u$. Thus, the Kronecker delta $\delta _{v,s}$ selects only one term in the sum with $v=s$ and we get an expression for $\beta(u,v)$ as,
\begin{equation}
\label{betaFinal}
\beta(u,v) = (m-u)(n-v)	\:	.
\end{equation} 
The total number of distinct projected grid element centers for a given $\chi_{k}$ can be calculated from Eq. (\ref{distinctPoints}),
\begin{equation}
\label{distinctPoints2}
{N}(u,v) \equiv {N}(\chi_{k}) = nm - (m-u)(n-v)	= mv + nu - uv	\:	.	
\end{equation}
This can be made more suggestive by writing in a form which allows comparison with $N_{0}$,
\begin{equation}
\label{distinctPoints3}
{N}(\chi_{k}) = N_{0} \left( \frac{v}{n} + \frac{u}{m} - \frac{uv}{N_{0}} \right) 	\:	.
\end{equation}
Each of these $N(\chi_{k})$ distinct points may have a finite number of other grid elements exactly eclipsed behind these distinct elements. The special cases of $\theta = 0$ and $\theta = \frac{\pi}{2}$ can be easily recovered for which $N$ is $m$ and $n$, respectively. We can see that ${N}(\chi_{k}) < N_{0}$ since the factor in the parentheses in Eq. (\ref{distinctPoints3}) is less than unity. This is expected because some elements will have an overlapping projection and the total number of distinct projections will decrease, as is the essence of the full-eclipse case. For these degenerate angles $\chi_{k}$, the total length of projection due to the grid elements will be denoted by $L_{P}(\chi_{k})$ and given by,
\begin{equation}
\label{degenerate_Lp}
L_{P} (\chi_{k}) = 2 N(\chi_{k}) a_{0} \: 	\:	,
\end{equation}
since each of the $N(\chi_{k})$ points contributes $2a_{0}$ to the projection length of the chord.
As we change $\theta$ in the range $\left\lbrace\chi_{k} \pm \Delta \chi_{k}\right\rbrace$, the full-eclipse degeneracy breaks and the grid elements behind every distinct element will begin to separate until all the $N_{0} = nm$ elements have completely distinct projections. In this angular range, there will be a finite overlap between the projections of the grid elements which were degenerate at $\chi_{k}$ and this would happen for all such $N(\chi_{k})$ points. This angular width and the corresponding projection length due to the grid elements will, in general, depend on the number of overlapping elements with the concerned distinct element at at $\chi_{k}$. Using Eq. (\ref{betat}), we see that ``\textit{Row - $t$}" has,
\begin{equation}
\label{rowt_distinct}
\Omega_{t}(\chi_{k}) = m - \beta _{t} (u,v) = m - (m-u) \sum_{k = 0}^{t-1} \delta_{v,k} 	\:	\:	,
\end{equation}
number of distinct element centers, which can be labeled as $\left\lbrace 1, 2, 3, .., \Omega_{t}(\chi_{k}) \right\rbrace$ from the right. Thus, each of the $N(\chi_{k})$ distinct element centers at $\theta = \chi_{k}$ can be identified with a coordinate $(t,\omega(t))$ with $1 \leq t \leq n$ and $1 \leq \omega(t) \leq \Omega_{t}(\chi_{k})$ such that $t$ labels the row and $\omega(t)$ the column. In the full-eclipse scenario, it can readily be seen that for the distinct grid element $(t,\omega(t))$ will be overlapping with a total of  $J(t,\omega(t),\chi_{k})$ number of elements at $\chi_{k}$, given by,
\begin{equation}
\label{Bt}
J(t,\omega(t),\chi_{k}) = \textsf{min}\left( \lfloor \frac{m-\omega(t)}{u}\rfloor , \lfloor\frac{n-t}{v}\rfloor \right) \: + 1	\:	\:	,
\end{equation}
where $\lfloor.\rfloor$ is the floor function\cite{floor} (The additional unity comes from the element itself). We now evaluate an expression for $\Delta \chi_{k}$, the angular width around $\chi_{k}$ needed to be swept to project all $J(t,\omega(t),\chi_{k})$ points separately. It can be seen that this width will only depend on $(u,v)$ and $a_{0}$, and not on $J(t,\omega(t),\chi_{k})$ since the projection is done using a plane wave-front. We define the distance between the centers of two adjacent, overlapping points at $\chi_{k}$ as $\mu$ which has the form,
\begin{equation}
\label{mu}
\mu = \sqrt{u^2 x^{2}_{0} + v^2 y^{2}_{0}} \: \: .
\end{equation}
At $\theta = \chi_{k} \pm \Delta\chi_{k}$, the projection of all the $J(t,\omega(t),\chi_{k})$ overlapping points, for each of the $N(\chi_{k})$ points will be just separated and tangential to each other, such that,
\begin{equation}
\label{deltachi}
\Delta\chi_{k}  = \sin^{-1}\left( \frac{2a_{0}}{\mu} \right)	\:	\:	.
\end{equation}
At an intermediate angle $\theta \in \left[\chi_{k} - \Delta\chi_{k} , \chi_{k} + \Delta\chi_{k} \right]$, there will be some finite overlap between grid elements that were degenerate at $\chi_{k}$, for each of the $N(\chi_{k})$ points. The projected chord length of the grid elements due to the overlapping points associated with the $(t,\omega(t))$ point can be written as,
\begin{equation}
\label{Lpt}
L_{P,t}(\theta,\chi_{k}) = \left[ J(t,\omega(t),\chi_{k})  - 1 \right] \mu \sin\vert\theta - \chi_{k}\vert + 2a_{0}	\:	\:	,
\end{equation}
where the subscript $``t"$ denotes that the point $(t,\omega(t))$ is under consideration and the argument $\chi_{k}$ refers that the angle is being varied around $\theta = \chi_{k}$ in the range $ \left[\chi_{k} - \Delta\chi_{k} , \chi_{k} + \Delta\chi_{k} \right]$. For all $N(\chi_{k})$ such sets of overlapping points, the total projected grid element length can be obtained by summing over both $t$ and $\omega(t)$ in Eq. (\ref{Lpt}),
\begin{eqnarray}
\label{Lpchi}
L_{P}(\theta, \chi_{k}) &=& \left[ \sum_{t = 1}^{n} \sum_{\omega(t) = 1}^{\Omega(t)} \left( J(t,\omega(t),\chi_{k}) - 1 \right) \right] \mu \sin\vert \theta - \chi_{k}\vert \nonumber \\ &+&   \left( 2a_{0}  \right) \left( \sum_{t = 1}^{n} \sum_{\omega(t) = 1}^{\Omega(t)} (1) \right) \:	\:	.	\nonumber\\	
\end{eqnarray}
It can be identified that the second double summation in Eq. (\ref{Lpchi}) is the total number of distinct grid elements at $\theta = \chi_{k}$ and therefore, Eq. (\ref{Lpchi}) may be written as,
\begin{eqnarray}
\label{Lpchi_final}
L_{P}(\theta, \chi_{k}) &=& \left[ \sum_{t = 1}^{n} \sum_{\omega(t) = 1}^{\Omega(t)} \left( J(t,\omega(t),\chi_{k}) - 1 \right) \right] \mu \sin\vert \theta - \chi_{k}\vert \nonumber  \\ &+&    2a_{0} N(\chi_{k}) 	\:	\:	.
\end{eqnarray}
We now define a quantity called the ``coverage fraction" which is a dimensionless number, giving us the fraction of the total projected chord length of the grid ${l(\theta)}$ which is covered by the projection of the grid elements. On the circular screen, this will correspond to the fraction of the length over which the photo-detector will detect a lower intensity than the background illumination due to the source plane wave-front which passes through the grid. This quantity $\lambda(\theta)$ can be measured and inferences about the grid can be made from these observations, as will be elaborated in the next section. It can be defined as follows,
\begin{equation}
\label{lambda_def}
\lambda (\theta) = \frac{L_{P}(\theta)}{{l(\theta)}}	\:	\:	,
\end{equation}
where ${l(\theta)}$ can be used from Eq. (\ref{ltheta_chord}) and $L_{P}(\theta)$ from Eq. (\ref{Lpchi_final}) or (\ref{L_P_nondegenerate}), depending on whether  $\theta \in \left[\chi_{k} - \Delta\chi_{k} , \chi_{k} + \Delta\chi_{k} \right]$ or not, respectively. Whenever $\theta \in \left[\chi_{k} - \Delta\chi_{k} , \chi_{k} + \Delta\chi_{k} \right]$, the coverage fraction would be lower, owing to the overlap in the grid projection. Thus, for each $\theta = \chi_{k}$, there would be a finite angular width in which the coverage fraction would decrease from its non-degenerate value.
In the next section, we present an application of our model of the grid projection to determine the unknown grid parameters $n,m,x_{0}, y_{0}$ and $a_{0}$ of a grid whose spatial extent is known. 
\section{Grid Parameter Determination - The Two Angle Approach} \label{two_angle}
With a mathematical model of the grid projection ready, we present a formulation to determine unknown parameters of the grid whose spatial extent is known. This will be done by measuring of the first two angles at which degeneracy exists in the projection. Consider a situation in which we have a grid with a known extent in $x$ and $y$ directions i.e. $L_{x}$ and $L_{y}$ are known. The grid parameters $n,m , x_{0}, y_{0}$ and $a_{0}$ are unknown and we wish to find these, by measurement of the two smallest non-zero angles $\chi_{1}$ and $\chi_{2}$ at which there is degeneracy in the grid projection. 
Physically, this can be achieved by continually changing the angle of the light beam (as described in section \ref{grid}) from $\theta = 0$ to increasing values and measuring the projected length of the grid elements on the screen. $\chi_{1}$ is the first or smallest full-eclipse angle at which there is a fall of the grid element length projection and $\chi_{2}$ will be the next largest such angle. With the two angles known, we would be able to determine the grid parameters as demonstrated here. It is easily seen that these two angles will be such that they follow,
\label{phi12}
\begin{eqnarray}
\tan \chi_{1} = \frac{x_{0}}{L_{y}} 		\: ,\label{phi1}\\
\tan \chi _{2} = \frac{x_{0}}{L_{y} - y_{0}} 	\: . 	\label{phi2}
\end{eqnarray}
Once we have measured $\chi_{1}$, we can readily infer $x_{0}$ using Eq. (\ref{phi1}) since $L_{y}$ is known. We can then use Eq. (\ref{lxly}) and our knowledge of $L_{x}$ to determine the number of columns in the grid,
\begin{equation}
\label{mFind}
m = \frac{L_{x}}{L_{y} \tan \chi_{1}} + 1	\:	.
\end{equation}
The measurement of the other angle $\chi_{2}$ can be then used to deduce $y_{0}$ using Eq. (\ref{phi2}) to get,
\begin{equation}
\label{y0Find}
y_{0} = L_{y} -  \frac{x_{0}}{\tan \chi_{2}}	\:	,
\end{equation}
which can be then used together with Eq. (\ref{lxly}) to get the value of the number of rows $n$ in the grid,
\begin{equation}
\label{nFind}
n = \frac{\tan \chi_{2}}{\tan \chi_{2} - \tan \chi_{1}} + 1	\:	.
\end{equation}
To determine the grid element size $a_{0}$, we would need the value of angular width where the first degeneracy occurs \textit{i.e.} $2\Delta\chi_{1}$ which can be measured by the photo-detector. At $\chi_{1}$, we see that $\mu \equiv \mu_{1} = \left( {x^{2}_{0} + (n-1)^2 y^{2}_{0}}  \right)^{1/2}$ and $N(\chi_{1}) = m(n-1) + 1$. Substituting in the expression for $\Delta\chi_{1}$ at $\theta = \chi_{1}$ from Eq. (\ref{deltachi}), we get an expression for $a_{0}$ as,
\begin{equation}
\label{a0Find}
a_{0} = \frac{1}{2} \mu_{1} \sin(2\Delta\chi_{1})   \:	\:	.
\end{equation}
Thus, with the knowledge of the physical extension of the grid and measurement of the two smallest angles for the degenerate case, we are successfully able to deduce the grid parameters using a suitable application of the mathematical model developed in the previous sections.
\section{Data Extraction Applications to Modern Science} \label{applications}
In this section, we present a short discussion on possible applications of our model to different setups in modern science. The model, which uses the projections of a rectangular grid onto a circular screen, in essence provides us with a methodology for data extraction techniques. The occurrence of an ordered structure in physics and other branches of science is fairly common and individual problems can motivate specific adaptations of the model. A given grid structure in three dimensions can be thought of consisting of many two-dimensional slices of grids and our model can be applied to each of these slices separately, while also be used to estimate the spacing between adjacent slices. (Ref. section \ref{two_angle}). The model harnesses the symmetry in the structure, which in our case of a rectangular grid is a $C_{2}$ symmetry about the $x$ and $y$ axes. The total projection of the grid onto the screen $\widehat{l(\theta)}$ will be representative of this symmetry if $\theta$ is allowed to vary from $0$ to $2\pi$. A simple plot of $\widehat{l(\theta)}$ with $\theta$ will easily give us estimates of the spatial extent of the structure \textit{i.e.} $L_{x}$ and $L_{y}$ in our case. For grid arrangements having a different symmetry such as triangular, hexagonal or circular etc.; the symmetry properties\cite{symmetry, symmetry2} can be obtained by studying the polar plot of $\widehat{l(\theta)}$ for $\theta \in [0,2\pi]$. 
The application of the model to classical macroscopic systems can be thought which use the angular pattern of the grid projection to infer properties of the system. Characterization of periodically arranged mechanical systems (eg. \cite{sonic}) can be implemented using this model. Our model can also be used to study the physical properties of a material, in particular its elasticity and response to temperature change. A known grid whose angular projections are known can be placed on a given material, which is then subject to changes. An external stress would tend to produce a strain in the material, altering the spacing between the grid points and may even alter the size of each grid point. A similar effect would be produced by the application of an external pressure. By measuring the changes in the grid projection pattern due to these external influences, one can determine elasticity\cite{elastic} properties of the material, such as its Young's Modulus\cite{srivastava} and Bulk Modulus\cite{srivastava}. This method can be easily extended to obtain the temperature expansion coefficient\cite{srivastava} of the material since we expect changes in its physical dimensions due to a temperature change. Exact implementation of these techniques using the grid projection would need a detailed construction of the setup and a proper analysis of all underlying assumptions, if any - such as uniformity of properties, accuracy of the projection etc. 
We also motivate the use of this model in astrophysics, where it can be used to measure the curvature of space-time and for the detection of gravitational waves. The model relies on changes in the physical extent of the grid due to variations in the local space-time, a characteristic common to ongoing large interferometric projects in this field like \textit{LIGO} and \textit{LISA}\cite{LISA}. 
Application to signal encoding can also be developed, whereby the grid pattern can be the encoded signal which can be transmitted at the speed of light to a receiver. The key to the decryption are the grid parameters which are only known to the sender and receiver. The information is contained in the particular angle at which the projection is transmitted. Such a methodology can be used for high-speed secure data transfer.
One of the areas of modern physics which deals with system having a regular structure is crystallography. Our method can serve as an addition to standard \textit{X-Ray Diffraction}\cite{diffraction1, diffraction2, diffraction3} and \textit{Neutron Diffraction} \cite{srivastava} techniques to determine the underlying crystal structure. Using our model, we can infer the orientation of the lattice planes and also estimate the spacing between adjacent parallel planes. Proper extensions of this work can also be used to characterize a Bravais Lattice completely by measuring all six of its lattice constants. Further thoughts on applications in crystallography can deal with measurement of the material density and detection and analysis of lattice defects\cite{srivastava}.
We emphasize that angular projection is an important and useful methodology to study the properties of a system since it is easy to implement and helps reducing the degrees of freedom of the data. In particular, the projection of a two-dimensional grid can be taken onto a one-dimensional screen, thereby reducing a degree of freedom to work with. Our work should be seen as a framework for a data extraction model which can be adapted suitably for different physical systems to appropriately extract details of the underlying structure by measuring its angular projection. In addition to data extraction uses, undergraduate students can pursue this as an engaging exercise to understand interesting aspects of the projection of a regular grid.
\section{Conclusion} \label{conclude}
We have presented a simple mathematical model for the projection of a rectangular grid onto a circular screen and discussed various applications to physical systems as data extraction techniques. Regular grids are a common geometric arrangement and our work tries to model the angular projection of the grid and relates the observed trend of physical observable quantities like the projected length, the number of projected points etc. with the grid parameters. We take into account the non-distinctness in the projection due to the overlap of projected grid elements in certain angular ranges, and give a sound methodology to deal with such cases leading to interesting trends for the coverage fraction. We presented an application of the model to determine the parameters of an unknown grid using the measurement of the first two angles at which degeneracy occurs.
Our model can be seen to have relevance and uses to formulate data extraction techniques. Applications of the model are motivated in different branches of modern science such as macroscopic elasticity property determination, astrophysics, crystallography and signal encoding.
Apart from such uses, such a model has a scholarary value to it. It can be of particular interest to undergraduate students who like working on simple, yet interesting, innovative problems to understand the structure of the solutions with sufficient rigour. 
\section*{Acknowledgments}
The author would like to thank Prof. Tashi Nautiyal Pant for proofreading an earlier version of the manuscript and Prof. Davinder Kaur \& Prof. Vipul Rastogi for their guidance. He is grateful to Megh Shah for helping with Figure \ref{gridsetup} and thankful to his colleagues, Satwik, Naveen and Aiman Khan for fruitful discussions on the topic.


\section*{References}
\begin{thebibliography}{10}
\bibitem{periodic} Ziman J M 1972 \textit{Principles of the Theory of Solids} Second Edition (Cambridge University Press)
\bibitem{kittle} Kittel C 1974 \textit{Introduction to Solid State Physics} Fifth Edition (Wiley Eastern, New Delhi)
\bibitem{bravais} Asokamani R 2006 \textit{Solid State Physics - Principles and Applications} (Anamaya Publishers, New Delhi)
\bibitem{hecht} Hecht E 2006 \textit{Optics} Fourth Edition (Pearson Education) 
\bibitem{photodetector} Paschotta R \textit{Photodetectors - RP Photonics Consulting GmbH} http://www.rp-photonics.com/photodetectors.html
\bibitem{kronecker} Weisstein E W, ``Kronecker Delta." \textit{From MathWorld--A Wolfram Web Resource.} http://mathworld.wolfram.com/KroneckerDelta.html 
\bibitem{floor} Weisstein E W ``Floor Function." \textit{From MathWorld--A Wolfram Web Resource.} 
http://mathworld.wolfram.com/FloorFunction.html 
\bibitem{symmetry} Ziman J M 1972 \textit{Principles of the Theory of Solids} Second Edition (Cambridge University Press) p 115
\bibitem{symmetry2} Ibach H and Luth 1991 \textit{Solid State Physics: An Introduction to Theory and Experiment} (Springer-Verlag Berlin Heidelberg) (Second Narosa Publishing House Reprint 1996)
\bibitem{sonic} Batra N K, Matic P and Everett R K 2002 \textit{IEEE Ultrasonics Symposium} p 547 - 50
\bibitem{elastic} Saxena B S, Gupta R C, Saxena P N 1996 \textit{Fundamentals of Solid State Physics} (Pragati Prakashan Meerut) Ch 4
\bibitem{srivastava} Srivastava C M and Srinivasan C 1997 \textit{Science of Engineering Materials} Second Edition
\bibitem{LISA} \textit{Laser Interferometer Space Antenna (LISA) Mission Concept} May 2009 ``LISA Project internal report number LISA-PRJ-RP-0001"
\bibitem{diffraction1} Saxena B S, Gupta R C, Saxena P N 1996 \textit{Fundamentals of Solid State Physics} (Pragati Prakashan Meerut) Ch 2
\bibitem{diffraction2} Ibach H and Lüth 1991 \textit{Solid State Physics: An Introduction to Theory and Experiment} (Springer-Verlag Berlin Heidelberg) (Second Narosa Publishing House Reprint 1996) Ch 3
\bibitem{diffraction3} Asokamani R 2006 \textit{Solid State Physics - Principles and Applications} (Anamaya Publishers, New Delhi)
\end{thebibliography}
\end{document}